# Chromatic dispersion and thermal coefficients of hygroscopic liquids: 5 glycols and glycerol

Daniel Jakubczyk[a], Gennadiy Derkachov[a], Kwasi Nyandey[a,b],
Sima Alikhanzadeh-Arani[a,†], Anastasiya Derkachova[a]

[a] Institute of Physics, Polish Academy of Sciences, al. Lotników 32/46, 02-668 Warsaw, Poland

[b] Laser and Fibre Optics Centre, Department of Physics, School of Physical Sciences, College of Agriculture and Natural Sciences, University of Cape Coast, Cape Coast, Ghana

[†] present address: Farhangian University, P.O. Box 14665-889, Tehran, Iran.

## Abstract
Chromatic dispersion and thermal coefficients of 6 hygroscopic liquids: ethylene glycol, diethylene glycol, triethylene glycol, tetraethylene glycol, propylene glycol (propane-1,2-diol), and glycerol were measured in the range from 390 to 1070 nm for temperatures from 1 to 45ºC. A modified Abbe refractometer was utilised. Special care was taken to avoid contamination of the liquids under the test with water and solid particles. The measurement uncertainties were analysed. It was noticed that (in the given range and within the available measurement accuracy) the dependence of the refractive indices on the wavelength and temperature could be considered independently. Thus, thermal coefficients were found for each wavelength used, and their weak dependence on the wavelength was recognised. Then the Sellmeier equation was fitted to the experimental results for each temperature.

## Introduction

Achieving ultimate accuracy in optical remote sensing and particle characterisation requires accurate values of refractive indices of characterised materials and host media for a given wavelength and temperature. Since we tackle such issues in our research (e.g. [1–3]), we've looked for the available refractive index data and usually found it insufficient for our purposes. Thus, we decided to build a dedicated setup to measure the refractive indices of liquids as a function of both wavelength and temperature. However, since we studied popular hygroscopic liquids, the results seem to be worth sharing. A specific application that may serve as a good example, is industrial dehydration of natural gas with glycols (most prominently triethylene glycol) [4]. The amount of absorbed water could be accurately assessed by refractive index measurement of the mixture, which calls, among others, also for accurate refractive index data of the pure liquids.

Accurate measurements of refractive indices – their chromatic dispersion and temperature coefficient in particular – of hygroscopic liquids, require special care to avoid contamination with water. A glimpse into e.g. Landolt-Börnstein database [5,6] reveals a large spread of results obtained by different authors over the past decades, which may be due precisely to pollution issues.

We shall briefly discuss some of the results that can be found in the literature on the ethylene glycol (EG) – a comparatively well investigated liquid, widely used as an engine coolant, antifreeze, de-icing agent, polymerisation precursor and desiccant. The results we discuss are presented in Fig. (2) together with the results of our experiments. Solid black dots represent different $n_{D20}$ measurements taken until late 1980s. It can be noticed that there are more outliers towards the lower values. The newer results of Tsierkezos [7] or Jimenez [8] – taken also at different temperatures (open stars and diamonds respectively), belong to the higher ones. The old (1916) but extensive results of Karvonnen on chromatic dispersion are in agreement with these. It seems to indicate that even before the molecular sieves came into everyday use, controlling the contaminating water content was possible with well-planned procedures. Our results are also in agreement with all the later mentioned. The contemporary measurements of chromatic dispersion in EG are rather sparse [9–11]. A fairly recent work by Sani and Dell'Oro [10,11] seems promising. An indirect method was utilised – the absorption (imaginary part of the refractive index) in EG was measured for a very wide range of wavelengths and Kramers–Kronig relation was invoked. However, neither the temperature, at which the dispersion curve was obtained (possibly 20ºC – room temperature), nor the purity of EG was stated. The older measurements of Voellemy [12] at 21.9 ºC and Timmermans et al. [13] at 15ºC are consistent with those, while contemporary measurement of Kozma et al. at 22ºC [9] clearly is not.

Most of the systematic measurements versus (either) wavelength or temperature are about century old. The measurements simultaneously dependent on wavelength and temperature are rare. In this work, we present such measurements we performed for 5 commercially available glycols and glycerol. We took extreme care not to contaminate them with water beyond the amount stated by the manufacturer, though we didn't use molecular sieves to avoid possible contamination with nanoparticles, which is crucial for our applications (light scattering). We measured dispersion of their refractive indices from 394 to 1070 nm for temperatures from 1 to 45ºC. We describe the setup and the procedures we used in detail.

## The refractometer setup

The experimental setup – see Fig. 1 – was based on a commercial Abbe refractometer (AR-4, Müller), which we modified to measure the chromatic dispersion of refractive index and its tem-perature dependence. We partly followed [14] (compare also e.g. [15]). So the compensator (2 Amici prisms) was set to maximum dispersion. In consequence, the index of refraction read from the scale had to be corrected with the formula (6) from [14], where we verified the prism material and apex angle by carefully calibrating the device with water. The light at a desired wavelength was provided through cylindrical lens from a monochromator (SPM 1, Carl Zeiss Jena [16]) with halogen lamp (H1 automotive bulb) illumination. Monochromator calibration was performed *in-situ* with a small grating spectrometer (USB4000, Ocean Optics; 1.4 nm resolution). We equipped the refractometer with a 14-bit digital camera (GC651MP, Smartek vision) looking through the refractometer's eyepiece with an additional camera objective ($f = 6$ mm, $f/1.6$, aberration correction including IR), so that both the bright and dark area (shadow resulting from the total internal reflection in the prism) with the crosshair superimposed and the scale were in the FoV simultaneously. In this way we could obtain sensible measurements in the spectral range from 390 to 1071 nm. The image of the bright and dark area was processed numerically using a Matlab program [17] written in the lab (clipping, background subtraction, vertical summation, smoothing) and the shadow edge brightness profile was obtained. The derivative of the profile exhibits a minimum corresponding to the inflection point on the brightness profile. The position of the shadow edge was identified with this point. The crosshair centre was determined by pointing at it in a magnified image at the beginning of the measurement series. Thus, the measurement consisted of adjusting the profile derivative minimum to the crosshair centre. The refractometer allows for measurements at different temperatures by circulating a liquid at a desired temperature through the prisms jacket. The temperature of the circulating cooling/heating liquid was maintained with in-lab built Peltier element-controlled heat exchanger with local stabilisation loop. A separate K-type thermocouple was placed directly next to the prisms-liquid contact surface to measure the temperature of the liquid accurately. It was measured with the calibrated CHY506R electronic thermometer (CHY Firemate Co.). Dry, filtered $N_2$ gas, obtained from liquid nitrogen, was flowed through the refractometer chassis and into the (plastic) protective housing in which the device was kept. The temperature of $N_2$ gas was stabilized by passing it through gas-liquid heat exchanger in a thermal bath stabilised down to ±0.2ºC. Depending on the temperature of the refractometer prisms, the temperature of the bath was set between –5 and 0ºC to keep it significantly lower than that of the prisms. The pipe leading the gas to the refractometer was thermally insulated, but the heat transfer was found significant there. Thus, the $N_2$ temperature was measured continuously at the entrance of the refractometer to ensure that during the experimental run it was stable down to ±1ºC. Depending on the temperature of the bath, it stabilised in the 1–6ºC range. In this way, we (i) stabilized the temperature of the refractometer body, which enabled good calibration of the device; (ii) prevented condensation of water vapour (as well as other vapours) on the optical surfaces in the device; (iii) minimized diffusion of atmospheric water into the measured liquid. The relative humidity (RH) measured in the enclosure was always below 24 % and the contact area of the sample in between the prisms with the enclosure atmosphere was ~10 $mm^2$ (in comparison to ~$10^3$ $mm^2$ in contact with glass). It is difficult to obtain a very low humidity in a fairly large plastic housing, because of relatively high residual water content in plastics (see e.g.: [18,19]). However, it was confirmed with a long-time measurement that the refractive index of a sample sitting between the prisms remains constant for several hours. An ample time (both for calibration and actual measurement) was always allowed for the temperature of the device to stabilise, where the primary condition was the temporal stability of the observed shadow line.

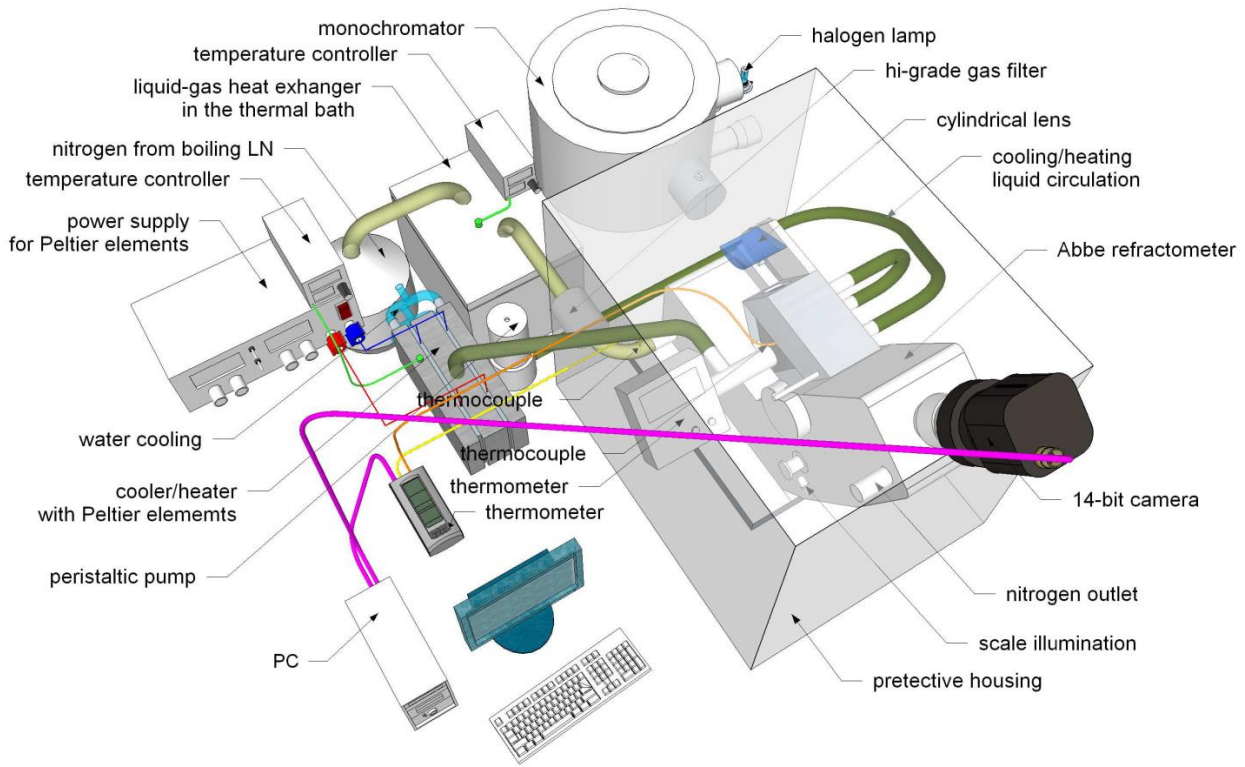

Fig. 1. Drawing of the measurement setup. Protective housing represented schematically.

Since for water there exist widely recognised systematic measurements of spectral dispersion versus temperature [20], the device was calibrated with water at a given temperature, before every measurement series. This fundamentally precluded calibration below 0ºC. Furthermore, in the case of water, for $T \to 0ºC$, $dn/dT \to 0$, which means that the calibration accuracy diminishes significantly towards 0ºC. Thus, for measurements at 1ºC the calibration was later augmented with data for EG we obtained. On the other hand, at elevated temperatures, the water between the prisms dried out faster, making calibration increasingly difficult. So, ~0.5 ml (an ample amount) of distilled water was slowly (to avoid forming bubbles) poured on the surface of the prism with a clean disposable syringe. The packages with disposable syringes were kept under vacuum to dry the syringe plastic as far as possible. The excess of water was squeezed out and removed on closing the top prism. The calibration was performed first of all at 589 nm (sodium D-line) and verified at 394 and 1071 nm. After the calibration, water was removed and the surface of prisms was carefully cleaned with a soft tissue and propanol. Then, the prisms were further dried with strong nitrogen gas flow (compressed $N_2$ of purity better than 99.8%). Extreme care must be taken, because the amount of liquid used for measurements is minute and even a small addition of water (or other substances) affects the measurements.

After the drying, ~0.5 ml of desired liquid was poured with a new disposable syringe as it was done with water and a series of measurements were taken at a desired temperature starting from the infrared towards the ultraviolet. Since the measurement series takes about 30 min, after the end, the measurements for the longest IR wavelengths were repeated, to ensure that there was no change in experimental conditions, e.g. due to mechanical creep in the apparatus. Furthermore, after the measurement of the sample, the prisms were cleaned and the calibration with water was rechecked, in order to exclude any systematic errors introduced by the creep in the apparatus taking place during the measurement of the sample. The unsealed bottles with the hygroscopic liquids were loosely recapped and stored under vacuum to ensure that they don't absorb atmospheric water. The lab was air-conditioned and the temperature of 22±1ºC was maintained, also to keep the humidity in the lab relatively low (~45%). However there was no stabilisation of hu-midity in the lab, and we observed some variation versus the season of the year. The measurement series for each temperature, were repeated several times, the outlying series were discarded and the remaining were averaged.

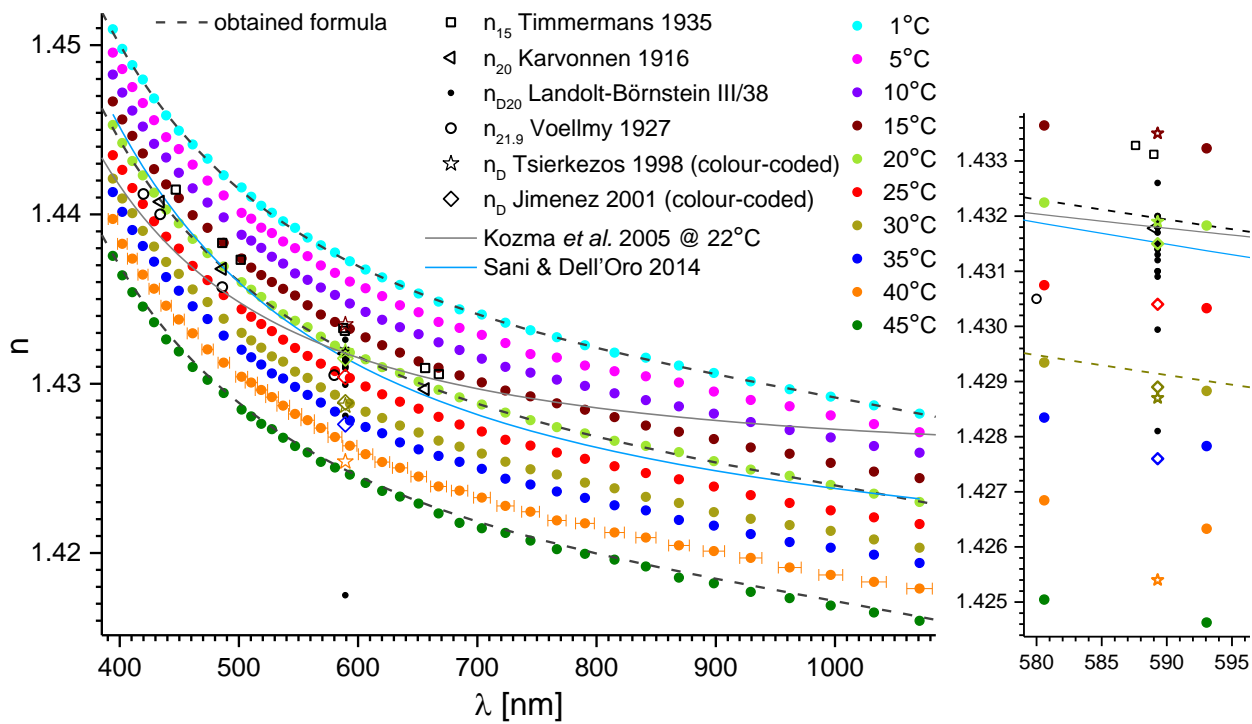

Fig. 2. Refractive index of ethylene glycol versus wavelength and temperature. Dashed lines represent fits with the obtained formula (Eqns. (1-3)) with parameters from Tabs. 1 and 2. The uncertainty of the wavelength is shown for clarity only for 40ºC in left panel. Markers pertaining to data from [7] and [8] are colour-coded according to coding of our data. Right panel: magnification of the region around the sodium D-line – most of the literature data falls in this region.

We sampled the following liquids: (i) ethylene glycol, anhydrous, 99.8% Sigma-Aldrich lot # STBG3967V; (ii) diethylene glycol, ≥99.0%, Sigma lot # BCBT4833; (iii) triethylene glycol, 99%, Alfa Aesar lot # 10198029; (iv) tetraethylene glycol, 99%, Alfa Aesar lot # M17E015; (v) propylene glycol (propane-1,2-diol), ≥99.5%, Sigma-Aldrich lot # MKC80613V; (vi) glycerol, BioUltra, anhydrous, ≥99.5%, Sigma lot # BCBS7814V.

## Data processing and accuracy considerations

After gathering the dataset $n(\lambda,T)$ – refractive indices for a set of (vacuum) wavelengths $\lambda$ and temperatures $T$ in the full available range (compare Fig (2)) – it was found that the dependence of $n$ on $\lambda$ and $T$ (in ºC) could be decomposed for each liquid under study, with $n(T, \lambda=\text{const})$ considered linear within our measurement accuracy (see Fig. 4):

$$n(\lambda,T) = n(\lambda, 20) + \frac{\mathrm{d}n(\lambda,T)}{\mathrm{d}T}(T-20) \ .$$

(1)

Thus, $\mathrm{d}n/\mathrm{d}T$ was found from a linear fit for each experimental $\lambda$ point. The mean relative standard error of $\mathrm{d}n/\mathrm{d}T$ is below 1% for all studied liquids. In inset in Fig. 4, two such fits at different $\lambda$ (central and peripheral) for EG are presented. Followingly, in Fig. 4 itself, we present $\mathrm{d}n/\mathrm{d}T(\lambda)$ for EG with the vertical error bars corresponding to $\mathrm{d}n/\mathrm{d}T$ standard error and the horizontal – to the estimated uncertainty of $\lambda$. Interestingly, $\mathrm{d}n/\mathrm{d}T$ displays a (weak) non-linear dependence on $\lambda$. A rational function, which is in line with the Sellmeier equation, was found to fit very well (COD=0.94 for EG):

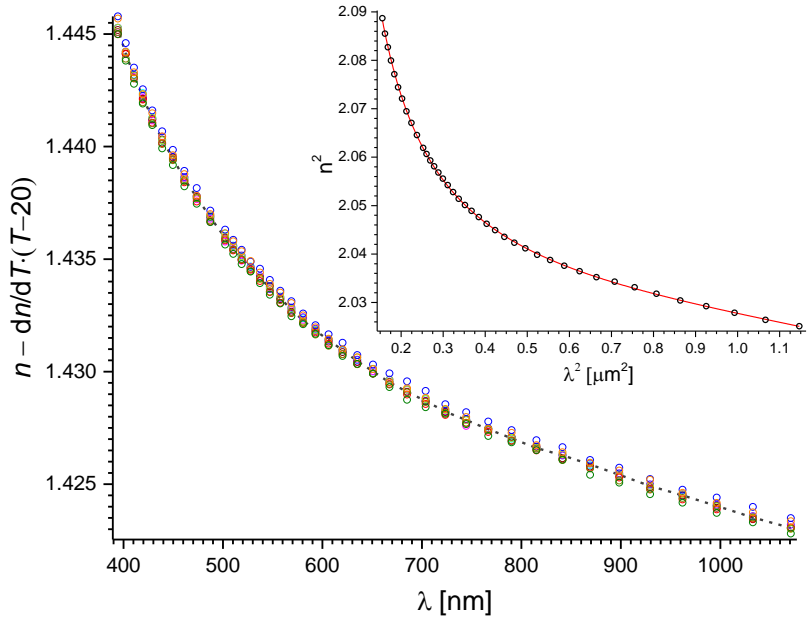

Fig. 3. The $n(\lambda)$ traces for all temperatures for EG were shifted to overlap the trace corresponding to 20ºC. The black dashed line shows the median of shifted traces. Inset: Sellmeier equation fitted (solid red line) to the obtained median points (black open circles).

$$\frac{dn(\lambda, T)}{dT} = A_T + \frac{B_T}{\lambda - C_T},$$

(2)

where $A_T$ constant is associated mainly with thermal expansivity (density change) of the liquid and $B_T$ and $C_T$ parameters have similar sense as in Sellmeier equation (see below). Then, using Eqn. 1, $n(\lambda)$ traces for all temperatures were shifted to overlap the trace corresponding to 20ºC (compare Fig. (3)). Median of all $n$ for each $\lambda$ was found and a two-pole Sellmeier equation

$$n^2(\lambda, 20) = A + \frac{B_{IR}\lambda^2}{\lambda^2 - C_{IR}} + \frac{B_{UV}\lambda^2}{\lambda^2 - C_{UV}}$$

(3)

was fitted, where $A$ accounts for the short-wavelength absorption contributions to $n$ at longer wavelengths, while $B_{IR}$ and $B_{UV}$ are absorption resonance strengths at wavelengths $C_{IR}^{1/2}$ and $C_{UV}^{1/2}$ respectively. The procedure was repeated for all studied liquids.

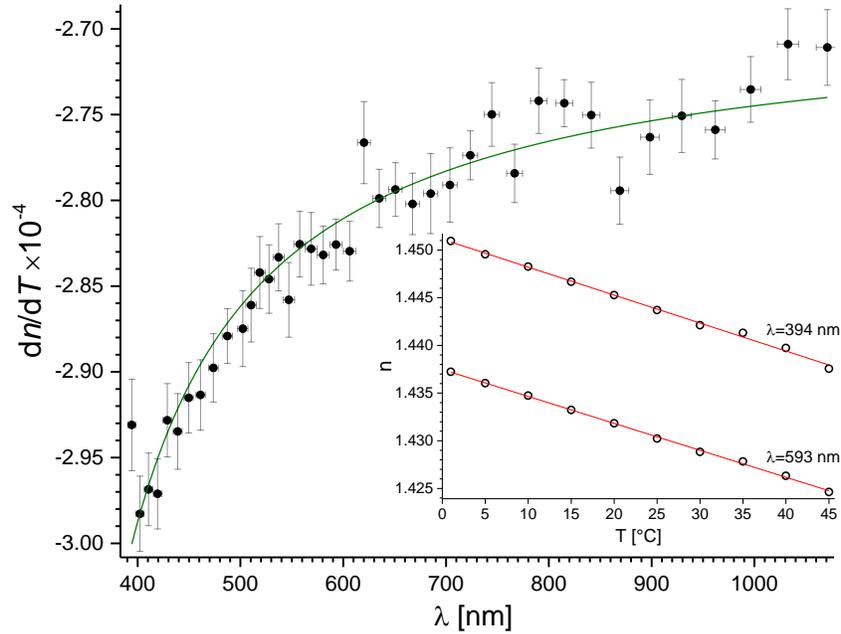

Fig. 4. Black solid circles: $dn/dT(\lambda)$ for EG obtained from linear $n(T)$ fits for each experimental $\lambda$ point. Vertical error bars correspond to $dn/dT$ standard error and the horizontal – to the estimated uncertainty of $\lambda$. Green solid line: rational function fit – Eqn. (2). Inset: open circles: $n(T)$; red solid lines: two linear fits at different $\lambda$ (central and peripheral).

### Error analysis

All the investigated samples were clear and colourless liquids, which indicates no significant absorption of light in the visible range. The literature data – where it is available (MEG [10,11], PG[21], glycerol[22]) – shows the absorption coefficient, in the range accessed in our experi-ment, to be at most of the order of 10-6. In consequence, the absorption of light in the samples has no impact on accuracy considerations.

The absolute precision of refractive index measurements in our setup (AR-4 refractometer equipped with additional lenses and a digital camera) was estimated as $\pm 1\times 10^{-4}$. The accuracy however was further influenced by mechanical hysteresis of the apparatus and its long term (mainly thermal) stability and was estimated as $\pm 3\times 10^{-4}$. So the vertical error bars in $n(\lambda,T)$ figures are of the size of the symbols – circles.

The maximal error of the wavelength determination was associated with the accuracy of monochromator-halogen lamp system calibration. The precision of wavelength setting is better than 1.5 nm. However, in order to achieve adequately bright illumination the slits were fairly wide-open which resulted in spectrally non-uniform illumination. Spectral profiles obtained with a multimode fibre with NA=0.22 (P600-1-SR, Ocean Optics) were not wider than 14 nm HWHM at 1071 nm and 3 nm HWHM at 394 nm. This led to the total calibration uncertainty of ~1%. The error bars for wavelength are shown for the trace corresponding to 40ºC as an example. Due to the character of the dispersion curve, the influence of these errors on the fitted Sellmeier curve is comparable to that introduced by refractive index uncertainty.

The accuracy of the calibrated CHY506R electronic thermometer with K-type thermocouple – traced to temperature standard – was estimated as ±0.2 K. Since a significant (up to 2K) temperature difference between prisms-liquid contact surface and the circulating liquid was observed, the temperature gradient across the prisms surface was checked with a thin calibrated (T-type) thermocouple (TT-T-40-SLE, by Omega, connected to CHY506R). It was found not greater than 0.4K, for highest temperature gradients in the setup (45ºC at the prisms, 2ºC at $N_2$ inlet to the refractometer, 22ºC ambient in the lab). So, finally, the accuracy of temperature measurements can be estimated as −0.3/+0.6K. Again, the horizontal error bars in Fig. d$n$/d$T(\lambda)$ (inset) would be of the size of the symbols.

As mentioned above, a significant error can be introduced into the measurement, if a hygroscopic sample is exposed to an ambient (humid) atmosphere, especially for longer periods. For instance, at 24% RH (as in the setup enclosure) the equilibrium water content in EG is ~9 wt% (see Fig. 18 in [23]). This would lead to a decrease of refractive index by ~0.01 (compare formula at Fig. 14 in [23]). It corresponds to 10% of whole difference between refractive indices of EG and water ($n_{EG} - n_{H2O}$). Similarly, the residual uncertainty of the refractive index due to the contamination with water for the EG lot that we used (see section "The refractometer setup") could be estimated as $\sim 2\times 10^{-4}$ (0.2% of $n_{EG} - n_{H2O}$), so the corresponding vertical error bars in $n(\lambda,T)$ figures would be below the size of the symbols – circles. In view of the above analysis, extreme care was taken to a avoid any prolonged contact of samples with the atmosphere or water-saturated containers – syringes, and the possible water intake during the experiment was carefully monitored as described in the previous sections.

As explained above, the measurements, in which the systematic errors were spotted by recalibration of the setup with water, were simply discarded.

The standard errors of Sellmeier equation coefficients and thermal coefficients (Tabs.1 and 2) are reported by the fitting procedure (simplex algorithm). Since, from the point of view of the optimization algorithm, both equations are over-parameterized, the errors tend to be significant, and the exact physical meaning of the values obtained is somewhat questionable. However, they are quite satisfactory from an engineering point of view.

## Results

The obtained Sellmeier equation coefficients and thermal coefficients, for $\lambda$ expressed in µm, are presented in Tabs. 1 and 2 respectively. All the obtained datasets (tables) are stored in Mendeley Data repository [24]. In Supplementary material (see below) we also present corresponding $n(\lambda,T)$ graphs for all studied liquids, comprising also relevant data from the literature. In case of less popular liquids, the chromatic dispersion data was not available for comparison.

| liquid | Sellmeier equation coefficients | | | | |
|---|---|---|---|---|---|
| | $A$ | $B_{IR}$ | $C_{IR}$ | $B_{UV}$ | $C_{UV}$ |
| ethylene glycol | 0.6 ± 0.8 | 0.0076 ± 0.0051 | 2.5 ± 0.7 | 1.4 ± 0.8 | 0.007 ± 0.004 |
| diethylene glycol | 1.6 ± 0.1 | 0.03 ± 0.05 | 6 ± 7 | 0.5 ± 0.1 | 0.02 ± 0.003 |
| triethylene glycol | 1.24 ± 0.25 | 0.008 ± 0.009 | 3.1 ± 1.8 | 0.85 ± 0.25 | 0.012 ± 0.003 |
| tetraethylene glycol | 1.48 ± 0.11 | 0.0035 ± 0.0032 | 2.25 ± 0.81 | 0.62 ± 0.11 | 0.0169 ± 0.0025 |
| propylene glycol | 1.15 ± 0.28 | 0.008 ± 0.006 | 2.75 ± 0.95 | 0.9 ± 0.3 | 0.011 ± 0.003 |
| glycerol | 1.61 ± 0.12 | 0.06 ± 0.09 | 8 ± 9 | 0.54 ± 0.12 | 0.018 ± 0.003 |

Tab. 1 Sellmeier equation coefficients for $\lambda$ in µm for 5 glycols and glycerol, found from the presented experiments. Uncertainties represent standard errors.

| liquid | thermal coefficients | | |
|---|---|---|---|
| | $A_T$ | $B_T$ | $C_T$ |
| ethylene glycol | $-2.643 \times 10^{-4} \pm 3 \times 10^{-7}$ | $-7.5 \times 10^{-6} \pm 1 \times 10^{-7}$ | 0.17 ± 0.01 |
| diethylene glycol | $-3.132 \times 10^{-4} \pm 2 \times 10^{-7}$ | $-5.61 \times 10^{-6} \pm 6 \times 10^{-8}$ | 0.245 ± 0.011 |
| triethylene glycol | $-3.122 \times 10^{-4} \pm 3 \times 10^{-7}$ | $-6.3 \times 10^{-6} \pm 2 \times 10^{-7}$ | 0.22 ± 0.03 |
| tetraethylene glycol | $-3.570 \times 10^{-4} \pm 2 \times 10^{-7}$ | $-6.51 \times 10^{-6} \pm 6 \times 10^{-8}$ | 0.235 ± 0.009 |
| propylene glycol | $-3.027 \times 10^{-4} \pm 3 \times 10^{-7}$ | $-1.2 \times 10^{-6} \pm 2 \times 10^{-7}$ | 0.72 ± 0.01 |
| glycerol | $-2.395 \times 10^{-4} \pm 5 \times 10^{-7}$ | $-6.2 \times 10^{-6} \pm 2 \times 10^{-7}$ | 0.18 ± 0.01 |

Tab. 2 Thermal coefficients for $\lambda$ in µm for 5 glycols and glycerol found from the presented experiments. Uncertainties represent standard errors.

## Acknowledgements

This research was funded in whole or in part by National Science Centre, Poland, grant 2021/41/B/ST3/00069. For the purpose of Open Access, the author has applied a CC-BY public copyright licence to any Author Accepted Manuscript (AAM) version arising from this submission

# Ethylene glycol

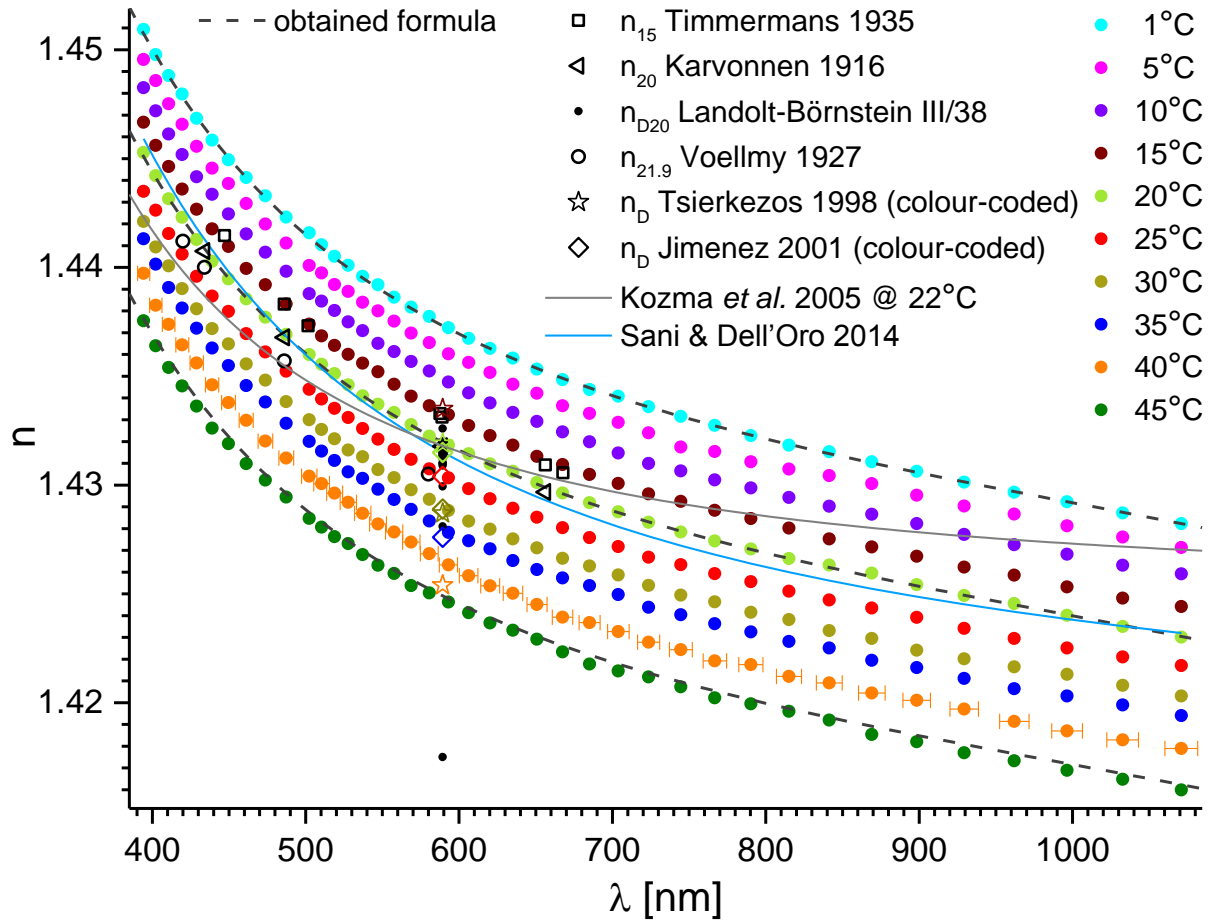

Refractive index of ethylene glycol versus wavelength and temperature. Dashed lines represent fits with the obtained formula (Eqns. (1-3) of the manuscript) with parameters from Tabs. 1 and 2 therein. The uncertainty of the wavelength is shown for clarity only for 40ºC. Essential data from Landolt-Börnstein database is also presented (black open squares; black open triangles; black dots; black open circles; brown, green, dark yellow, and orange open stars; green, red, dark yellow and blue open diamonds; grey and blue lines). Newer or more exhaustive data referenced directly.

# Diethylene glycol

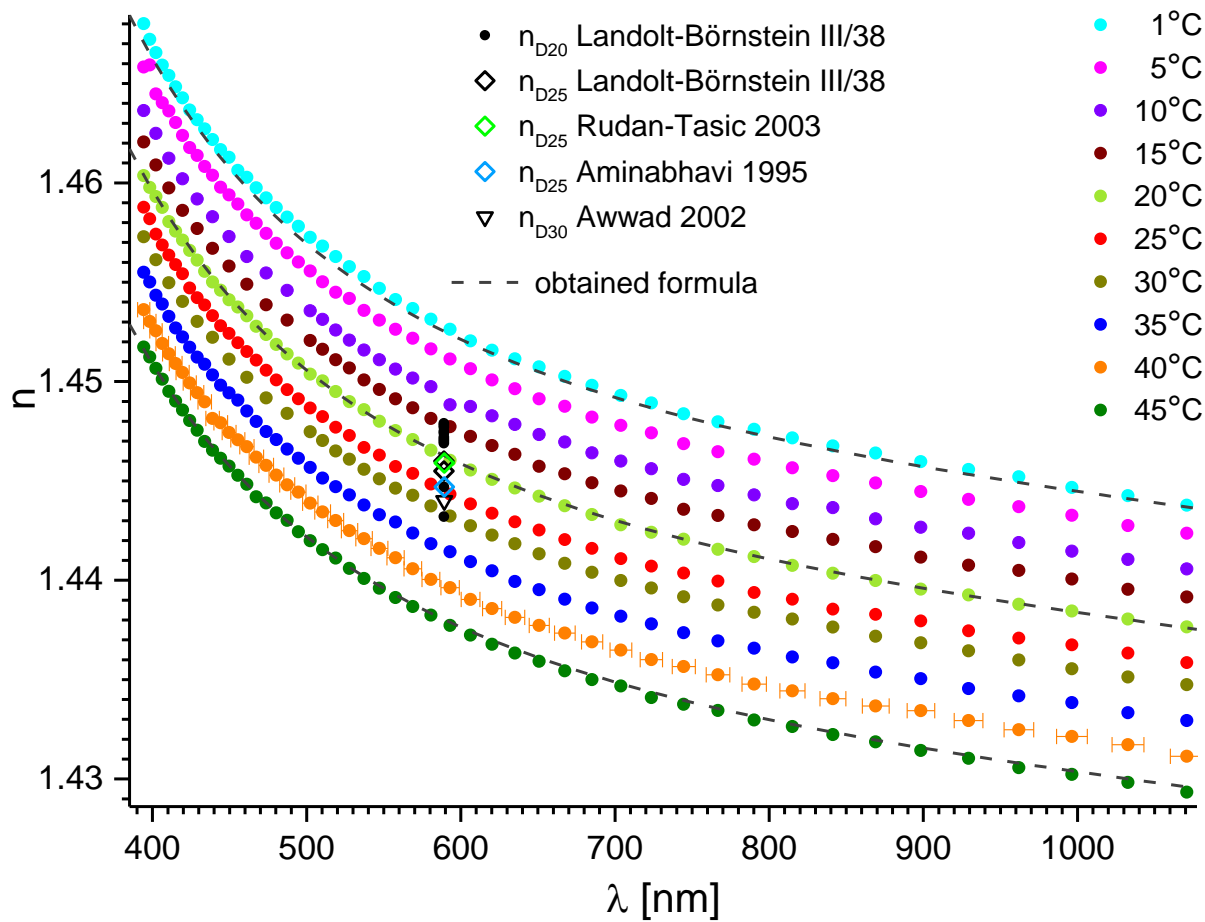

Refractive index of diethylene glycol versus wavelength and temperature. Dashed lines represent fits with the obtained formula (Eqns. (1-3) of the manuscript) with parameters from Tabs. 1 and 2 therein. The uncertainty of the wavelength is shown for clarity only for 40ºC. Essential data from Landolt-Börnstein database is also presented (black dots, black, green and blue open diamonds, black open triangle). Newer data referenced directly.

# Triethylene glycol

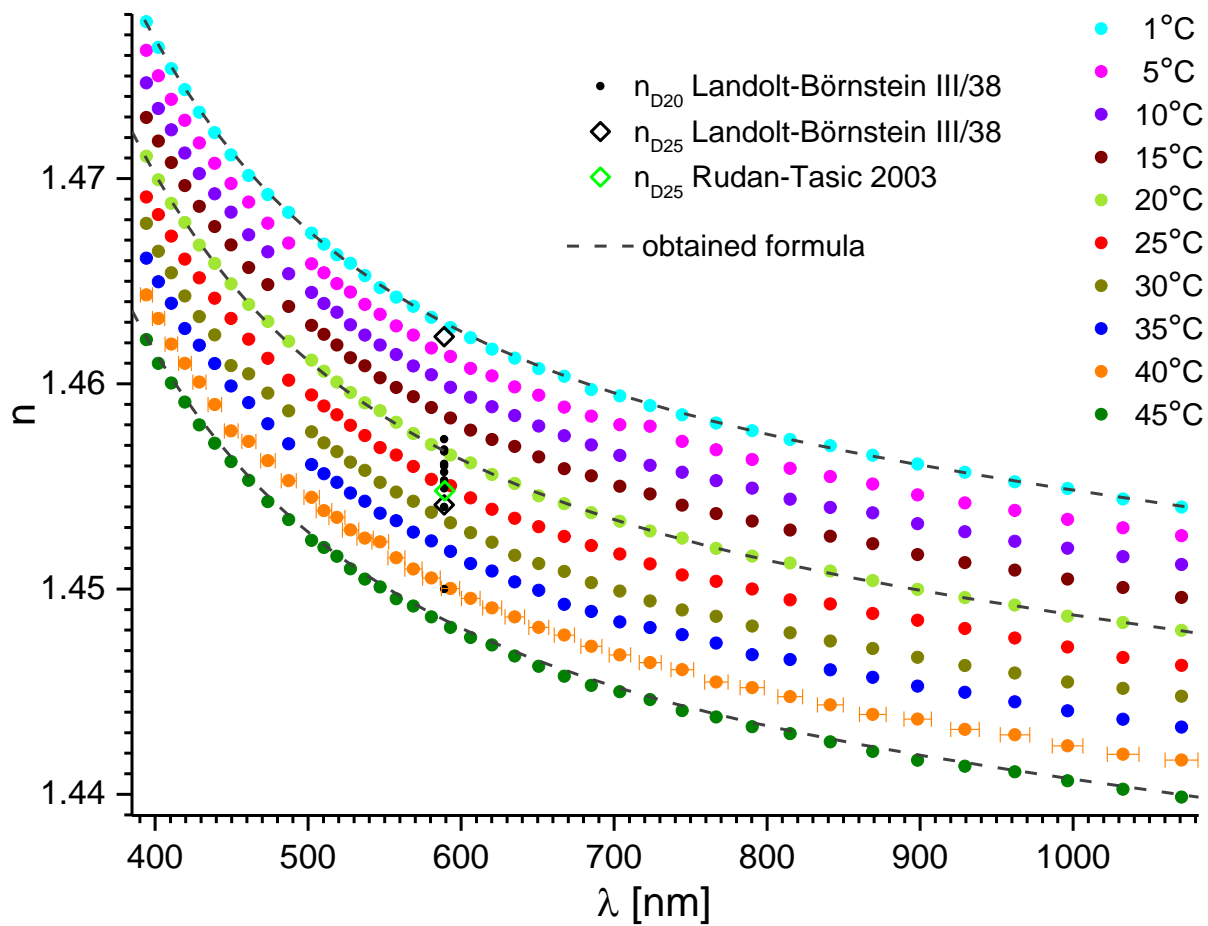

Refractive index of triethylene glycol versus wavelength and temperature. Dashed lines represent fits with the obtained formula (Eqns. (1-3) of the manuscript) with parameters from Tabs. 1 and 2 therein. The uncertainty of the wavelength is shown for clarity only for 40ºC. Essential data from Landolt-Börnstein database is also presented (black dots, black and green open diamonds). Newer data referenced directly.

# Tetraethylene glycol

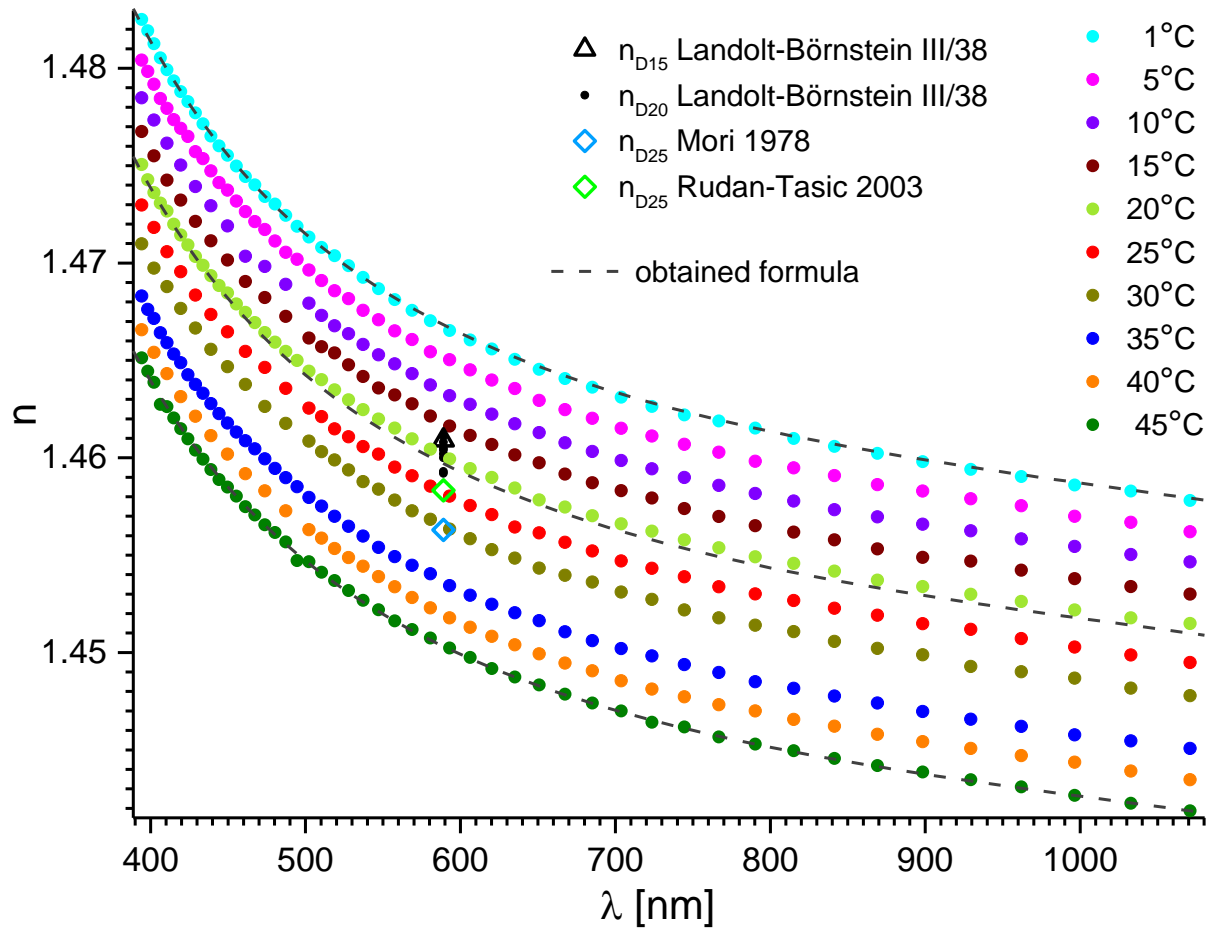

Refractive index of tetraethylene glycol versus wavelength and temperature. Dashed lines represent fits with the obtained formula (Eqns. (1-3) of the manuscript) with parameters from Tabs. 1 and 2 therein. The uncertainty of the wavelength is shown for clarity only for 40ºC. Essential data from Landolt-Börnstein database is also presented (black open triangles, black dots, green and blue open diamonds). Newer data referenced directly.

# Propylene glycol

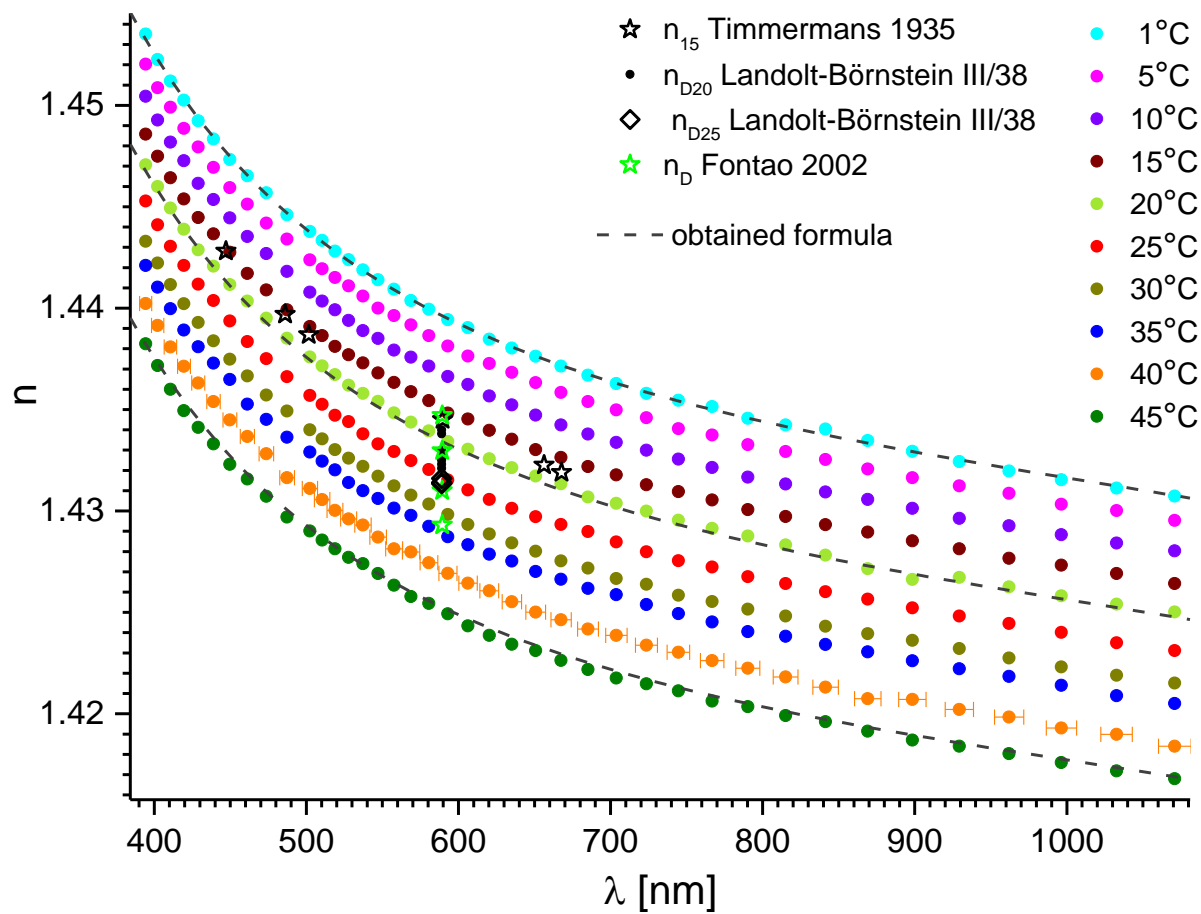

Refractive index of propylene glycol versus wavelength and temperature. Dashed lines represent fits with the obtained formula (Eqns. (1-3) of the manuscript) with parameters from Tabs. 1 and 2 therein. The uncertainty of the wavelength is shown for clarity only for 40ºC. Essential data from Landolt-Börnstein database is also presented (black dots, black open diamonds, black and green open stars). Newer or more exhaustive data referenced directly.

# Glycerol

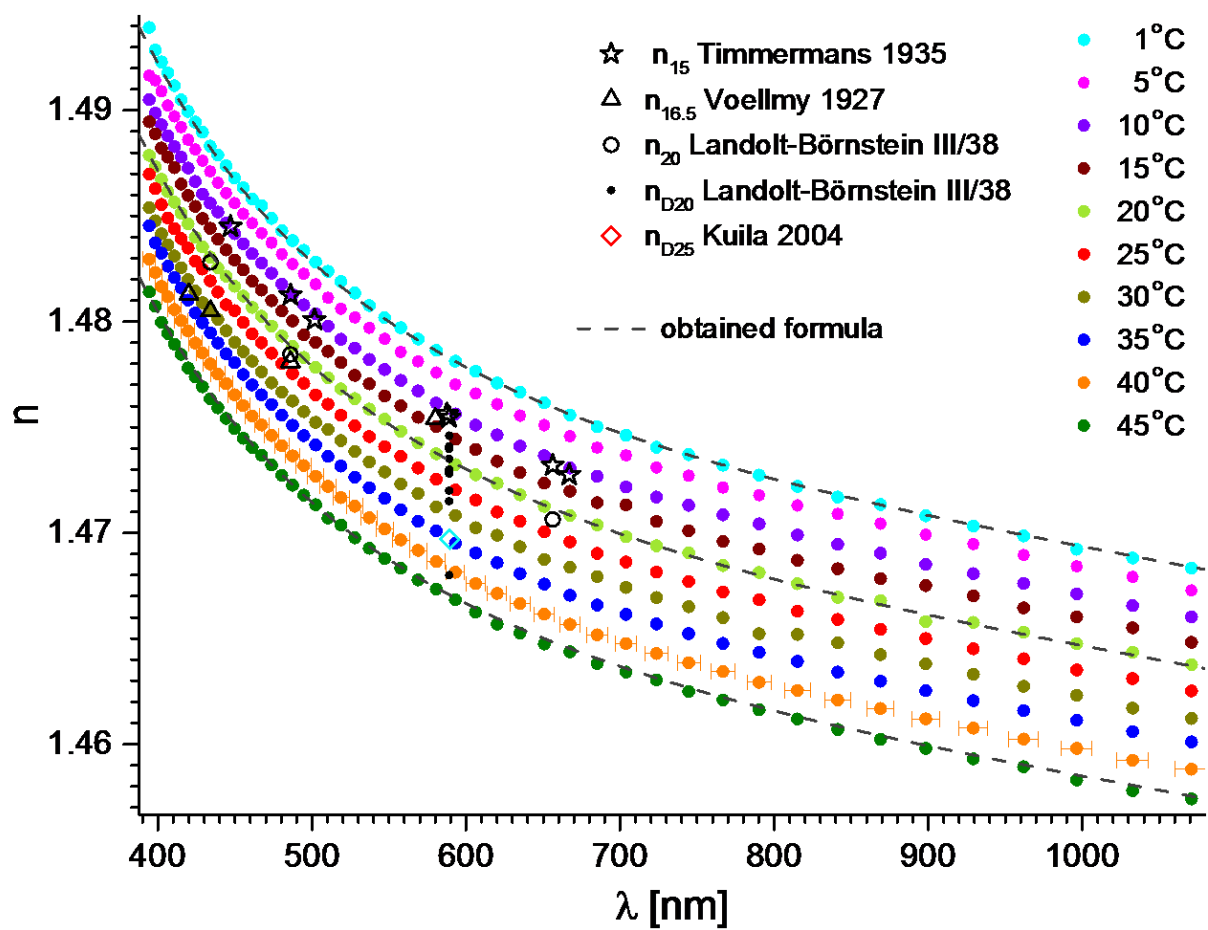

Refractive index of glycerol versus wavelength and temperature. Dashed lines represent fits with the obtained formula (Eqns. (1-3) of the manuscript) with parameters from Tabs. 1 and 2 therein. The uncertainty of the wavelength is shown for clarity only for 40°C. Essential data from Landolt-Börnstein database is also presented (black open stars, black open triangles, black open circles, black dots, red open diamond). Newer or more exhaustive data referenced directly.